\documentclass[prl,twocolumn,floats,aps,epsfig,nofootinbib,amssymb]{revtex4-1}
\pdfoutput=1
\usepackage{graphicx}
\usepackage{cancel}
\usepackage{amssymb}
\usepackage{textcomp}
\usepackage{amsmath}
\usepackage{bm}
\usepackage{times}
\usepackage{epsfig}
\usepackage{color}
\usepackage{graphics}
\usepackage{hyperref}
\usepackage{setspace}
\usepackage{slashed}
\usepackage{amsmath, amssymb}
\def\beq{\begin{equation}}
\def\eeq{\end{equation}}
\def\bea{\begin{eqnarray}}
\def\eea{\end{eqnarray}}

\hypersetup{
    pdfnewwindow=true,      
    colorlinks=true,       
    linkcolor=black,          
    citecolor=blue,        
    filecolor=blue,      
    urlcolor=blue           
}

\def\tr{\text{Tr}\,}

\begin{document}
\title{\Large  {\textit{\bf{The Higgs Mass via Type II Seesaw}}}}
\bigskip
\author{\large Pavel Fileviez P{\'e}rez}
\address{Particle and Astro-Particle Physics Division \\
Max-Planck Institute for Nuclear Physics (MPIK) \\
Saupfercheckweg 1, 69117 Heidelberg, Germany}
\author{\large Sogee Spinner}
\address{International School for Advanced Studies (SISSA) \\ and INFN, Via Bonomea 265, 34136 Trieste, Italy}
\date{\today}

\begin{abstract}
We show that the simplest supersymmetric scenario where a large Higgs mass can be attained at tree-level for any ratio of the Higgs vacuum 
expectation values corresponds to the case where the neutrino masses are generated through the type II seesaw mechanism. This allows a standard model-like 
Higgs with mass around 125 GeV without assuming a heavy spectrum for the stops.
We show that our results are consistent with the bounds coming from perturbativity 
up to the grand unified scale, discuss gauge coupling unification and possible signals at the Large Hadron Collider.
\end{abstract}
\maketitle
\newpage
\section{I. Introduction}
The recent discovery of a new particle~\cite{ATLAS,CMS} at the Large Hadron Collider (LHC) with properties similar to the Standard Model (SM) Higgs, is 
one of the most important and exciting discoveries to hit the particle physics world in the last several decades. Understandably, it has spurred a flurry of studies which attempt to understand the properties of the Higgs field. Supersymmetry (SUSY) provides a natural theoretical framework to study this issue since the minimal supersymmetric 
standard model (MSSM) predicts a tree-level Higgs mass no larger than the $Z$ mass. While this falls short of the recently measured 125 GeV particle, loop corrections from a heavy stop sector can complete the difference.  For a review on the predictions for the Higgs mass in the MSSM see Ref.~\cite{review-MSSM}.

Extending the MSSM can alleviate the tree-level Higgs mass bound. The next to minimal supersymmetric standard model (NMSSM) enlarges the MSSM Higgs content by an SM singlet and its coupling to the MSSM Higgs fields can increase the tree-level mass beyond $M_Z$. However, the new contribution is maximized for values of the ratio of the Higgs doublet vacuum expectation values (VEV), $\tan \beta$, close to unity, while the MSSM contribution is maximized for $\tan \beta$ far from unity. For a review of the NMSSM see~\cite{NMSSM}. Furthermore, significant enhancements require coupling values which are not perturbative to the grand unified theory (GUT) scale.
One can also extend by a hyperchargeless triplet~\cite{T1, T2} or a combination of the triplet and singlet~\cite{T3} but in both cases the new contribution to the Higgs mass occur for $\tan \beta$ close to unity and perturbativity is an issue, although gauge coupling unification considerations can solve the latter issue~\cite{Adjoint-MSSM}.

The singlet and the hyperchargeless triplet are two of the three possible fields that can introduce new quartic Higgs couplings. The third, triplets with hyperchage one, can contribute significantly to the Higgs mass independent of $\tan \beta$~\cite{T1,T2}. Coincidently, these fields also facilitate the type II seesaw mechanism for neutrino masses. 
In this article we show that such fields allow a large tree-level Higgs mass for any value of $\tan \beta$ while keeping all couplings perturbative to the GUT scale.  We also discuss the possibility of keeping the interesting MSSM prediction of gauge coupling unification. The most generic signatures at the LHC are briefly discussed.
 
This article is organized as follows. In Section II we survey the predictions 
for the Higgs mass in different supersymmetric models. In Section III we 
review the type II seesaw mechanism for neutrino masses and the properties 
of the Higgs sector. In Section IV we discuss gauge coupling unification and in Section V we highlight possible signals 
at the LHC. We summarize our main results in Section VI.

\section{II. The Higgs  Mass in the MSSM and Beyond}
The Higgs sector of the MSSM is composed of two Higgs doublets:
\begin{displaymath}
	\hat{H}_u = 
		\begin{pmatrix}
		\hat{H}_u^+
		\\
		\hat{H}_u^0
	\end{pmatrix} \sim (1,2,1/2),
\  \
\hat{H}_d = 
		\begin{pmatrix}
		\hat{H}_d^0
		\\
		\hat{H}_d^-
	\end{pmatrix} 
	\sim (1,2,-1/2),
\end{displaymath}	
resulting in three neutral physical Higgs fields: $h, H$ and $A$, and charged Higgs bosons $H^{\pm}$. 
In the decoupling limit, $M_A^2 >> M_Z^2$ ($M_Z$ is the mass of the $Z$ boson), the lightest CP-even Higgs, $h$, is 
SM-like with the tree-level upper bound:
$$M_h^2 \le M_Z^2 \cos^2 2 \beta, $$
requiring a large one-loop level contribution to be consistent with a Higgs interpretation of the new 125 GeV boson.

In the NMSSM, where a singlet superfields, $\hat{S}$, allows for the term $\lambda_H \hat{S} \hat{H}_u \hat{H}_d$, 
the upper bound changes to
$$M_h^2 \le M_Z^2 \cos^2 2 \beta \ + \  \frac{1}{2} \lambda^2_H v^2 \sin^2 2 \beta .$$
This effect is relevant for $\tan \beta$ close to unity
and for a large contribution to the Higgs mass $\lambda_H$ is not perturbative up to the GUT scale. An extra contribution can be gained  from introducing a real triplet, $\hat{\Sigma} \sim (1,3,0)$, which couples as $\lambda_\Sigma \hat{H}_u \hat{\Sigma} \hat{H}_d$ and increases the Higgs mass upper bound to 
$$M_h^2 \le M_Z^2 \cos^2 2 \beta \ + \  \frac{1}{2} \left( \lambda^2_H + \frac{1}{2} \lambda^2_\Sigma \right) \ v^2 \sin^2 2 \beta .$$
Unfortunately, as in the NMSSM this effect 
on the Higgs mass is only relevant for small of $\tan \beta$ and does not improve perturbativity. For the study 
of the Higgs sector in triplet extensions of the MSSM see Refs.~\cite{T1,T2,T3,Adjoint-MSSM,T4,T5}.
The main goal of this work is to investigate the simplest extension of the 
MSSM which can generate a large Higgs mass at tree-level for any value of $\tan \beta$ 
and to show how to keep the MSSM predictions. 

\section{III. Type II Seesaw and the Higgs Mass}
In order to implement the type II seesaw mechanism~\cite{TypeII} for neutrino masses 
in the MSSM, the Higgs content must be extended with two $SU(2)$ triplets:
\begin{equation}
	\hat{\Delta}_1 = 
		\begin{pmatrix}
		\frac{1}{\sqrt 2} \hat{\delta}_1^-
		&
		\hat{\delta}_1^{0}
		\\
		\hat{\delta}_1^{--}
		&
		-\frac{1}{\sqrt 2} \hat{\delta}_1^-
	\end{pmatrix} \sim (1,3, -1),
\end{equation}
\begin{equation}
	\hat \Delta_2 = 
	\begin{pmatrix}
		\frac{1}{\sqrt 2} \hat \delta_2^+
		&
		\hat \delta_2^{++}
		\\
		\hat \delta_2^0
		&
		-\frac{1}{\sqrt 2} \hat \delta_2^+
	\end{pmatrix} \sim (1,3, 1).
\end{equation}
In this case the relevant superpotential reads as
\begin{eqnarray}
	{\cal{W}}_{\rm{II}} & = &
		y_u \hat Q \hat H_u \hat u^c \ + \ y_d \hat Q \hat H_d \hat d^c
		\ + \ y_e \hat L \hat H_d \hat e^c
		\ - \  \mu \, \hat H_u \hat H_d 
		\nonumber \\
		&+&  \mu_\Delta \tr \hat \Delta_1 \hat \Delta_2
		\ + \  \lambda_1 \hat H_u \hat \Delta_1 \hat H_u
		\ + \  \lambda_2 \hat H_d  \hat \Delta_2 \hat H_d 
		\nonumber \\
		& + & f_\nu \hat L \hat \Delta_2 \hat L,
\end{eqnarray}
where the last term generates the neutrino mass matrix:
\begin{equation}
M_\nu = f_\nu \ \frac{v_2}{\sqrt{2}},
\end{equation}
where $v_2$ is the VEV of $\delta_2^0$. Notice that if $v_2 \sim 1$ GeV, $f_\nu$ must be small, 
about $10^{-9}$. The $\lambda_1$ and $\lambda_2$ terms in the superpotential allow for quartic Higgs couplings, which modify the upper bound on the lightest CP-even scalar tree-level mass to
\begin{equation}
	\label{ubII}
	M_h^2 \le \cos^2 {2 \beta} M_Z^2 \ + \ 2 \sin^4 \beta  \lambda_1^2 v^2 \ + \  2 \cos^4  \beta  \lambda_2^2 v^2.
\end{equation}
The new contributions can be sizable for any value of $\tan \beta$, but interestingly
the $\lambda_1$ and MSSM contributions both increases with  $\tan \beta$ allowing for constructive interfere. This is in contrast to the scenarios reviewed in the previous section where the additional quartic term (proportional to $\sin^2 {2 \beta}$) has the inverse $\tan \beta$ dependence of the MSSM contribution.
The corresponding soft SUSY breaking Lagrangian is
\begin{eqnarray}
	-\mathcal{L}_\text{Soft} & = &
		m_{H_u}^2 \left| H_u \right|^2 + m_{H_d}^2 \left| H_d \right|^2
		+m_{\Delta_1}^2 \left| \Delta_1 \right|^2+m_{\Delta_2}^2 \left| \Delta_2 \right|^2
		\nonumber \\
		&+ &
		\left(
			-b \, H_u H_d + b_\Delta \tr \Delta_1  \Delta_2
		\right.
			+a_{1} H_u \Delta_1 H_u
		\nonumber \\
		&+&
		\left.	
		a_{2} H_d \Delta_2 H_d 
		 + a_f  L  \Delta_2 L + \text{h.c.}
		\right) + ...,
\end{eqnarray}
where the remaining MSSM soft terms have been left out. The scalar potential coming from $D$-term 
is given by
\begin{eqnarray}
	V_D & = & \frac{1}{8} g_1^2 \left(\left|H_u\right|^2-\left|H_d\right|^2
		+ 2 \ \rm{Tr} \Delta^\dagger_2 \Delta_2 \  - \  2 \  \rm{Tr} \Delta_1^\dagger \Delta_1 \right)^2
		\nonumber \\ 
		& &
		+ \frac{1}{8} g_2^2  \sum_{a=1}^3 \left(H_u^\dagger  \sigma_a H_u + H_d^\dagger  \sigma_a H_d
		\right.
		\nonumber \\
		&&
		\left.
		+ 2 \  {\rm{Tr}} \Delta_1^\dagger \sigma_a \Delta_1  +  2 \ {\rm{Tr }} \Delta_2^\dagger \sigma_a \Delta_2  \right)^2.
\end{eqnarray}
The vacuum expectation values (VEVs) of the fields are defined to be
$$
	\left<H_u^0 \right> = \frac{v_u}{\sqrt 2}, \quad
	\left<H_d^0\right> = \frac{v_d}{\sqrt 2} ,\quad
\left< \delta^0_1 \right> = \frac{v_1}{\sqrt 2}, \quad
\left< \delta^0_2 \right> = \frac{v_2}{\sqrt 2}.
$$
The VEVs of the triplets break custodial symmetry and are therefore constrained by the $\rho$ parameter to be less than about 2-3 GeV~\cite{T5}. 
Therefore, working to zeroth order in $v_1$ and $v_2$ (the full minimization conditions are in the Appendix) the following minimization conditions can be derived:
\begin{align}
	0 & =
		m_{H_d}^2 + \mu^2 + \frac{1}{2} c_{2 \beta} M_Z^2 + t_\beta b + c_\beta^2 \lambda_2^2 v^2,
	\\
	0 & =
	m_{H_u}^2 + \mu^2 - \frac{1}{2} c_{2 \beta} M_Z^2 + \frac{1}{t_\beta} b
	+ s_\beta^2 \lambda_1^2 v^2,
\\
	\label{MC1}
		0 & =
			m_{\Delta_1}^2 + \mu_\Delta^2 + c_{2 \beta} M_Z^2
			+2 s_\beta^2 \lambda_1^2 v^2 +  \frac{v_2}{v_1} b_\Delta \nonumber
			\\
			& + \frac{v^2}{\sqrt 2 v_1}
			\left(
				c_\beta^2 \lambda_2 \mu_\Delta - s_{2 \beta} \lambda_1 \mu - a_1 s_\beta^2
			\right),
	\\
	\label{MC2}
		0 & =
			m_{\Delta_2}^2 + \mu_\Delta^2 - c_{2 \beta} M_Z^2
			+2 c_\beta^2 \lambda_2^2 v^2 \nonumber
			\\
			& +  \frac{v_1}{v_2} b_\Delta
			+\frac{v^2}{\sqrt 2 v_2}
			\left(
				a_2 c_\beta^2 + s_{2 \beta} \lambda_2 \mu - s_\beta^2 \lambda_1 \mu_\Delta
			\right).
\end{align}
Defining $v_1=v_\Delta \cos \gamma$ and $v_2=v_\Delta \sin \gamma$ we can solve for $v_\Delta$ and one finds
\begin{equation}
v_\Delta = \frac{v^2}{\sqrt{2} \cos \gamma M_1^2} \left( - c_\beta^2 \lambda_2 \mu_\Delta + s_{2 \beta} \lambda_1 \mu - a_1 s_\beta^2 \right),
\end{equation}
where
\begin{equation}
M_1^2= m_{\Delta_1}^2 \ + \  \mu_\Delta^2 \ + \  c_{2 \beta} M_Z^2 \ +\  2 s_\beta^2 \lambda_1^2 v^2 \ + \ \tan \gamma \ b_\Delta.
\end{equation}
Notice that if $M_1$ is around the TeV scale and the trilinear term $a_1$ is small, as in gauge mediation, one can get $v_\Delta \sim 1$ GeV, 
thereby  agreeing with the $\rho$ parameter constraints. It is important to mention that in the limit when $\lambda_1, \lambda_2, a_1, a_2 \to 0$ $B-L$ 
is an exact global symmetry. Therefore, one can say that the VEVs of the seesaw triplets are protected by the global $B-L$ symmetry.  
\begin{figure}[h] 
	\includegraphics[scale=0.7]{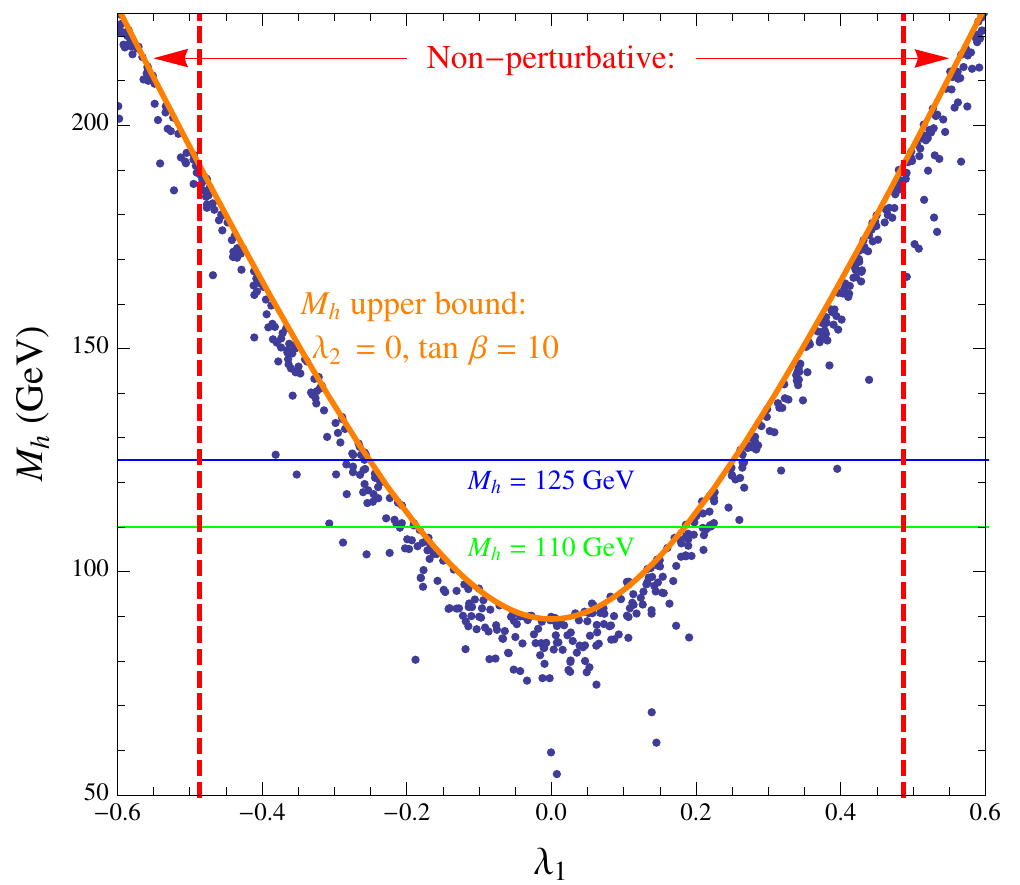}
\caption{The tree-level Higgs mass versus $\lambda_1$ with parameters scanned as described in the text. The orange curve represents the upper bound on the tree-level mass, Eq~(\ref{ubII}), with $\lambda_2 = 0$ and $\tan \beta =10$. To aid the eye,  horizontal lines at 110 GeV and 125 GeV are drawn in green and blue respectively. The area outside the vertical red dashed lines becomes non-perturbative before the GUT scale under the assumptions discussed in the next section.}
\label{mh.v.lambda1}
\end{figure}	
\begin{figure}[h] 
	\includegraphics[scale=0.7]{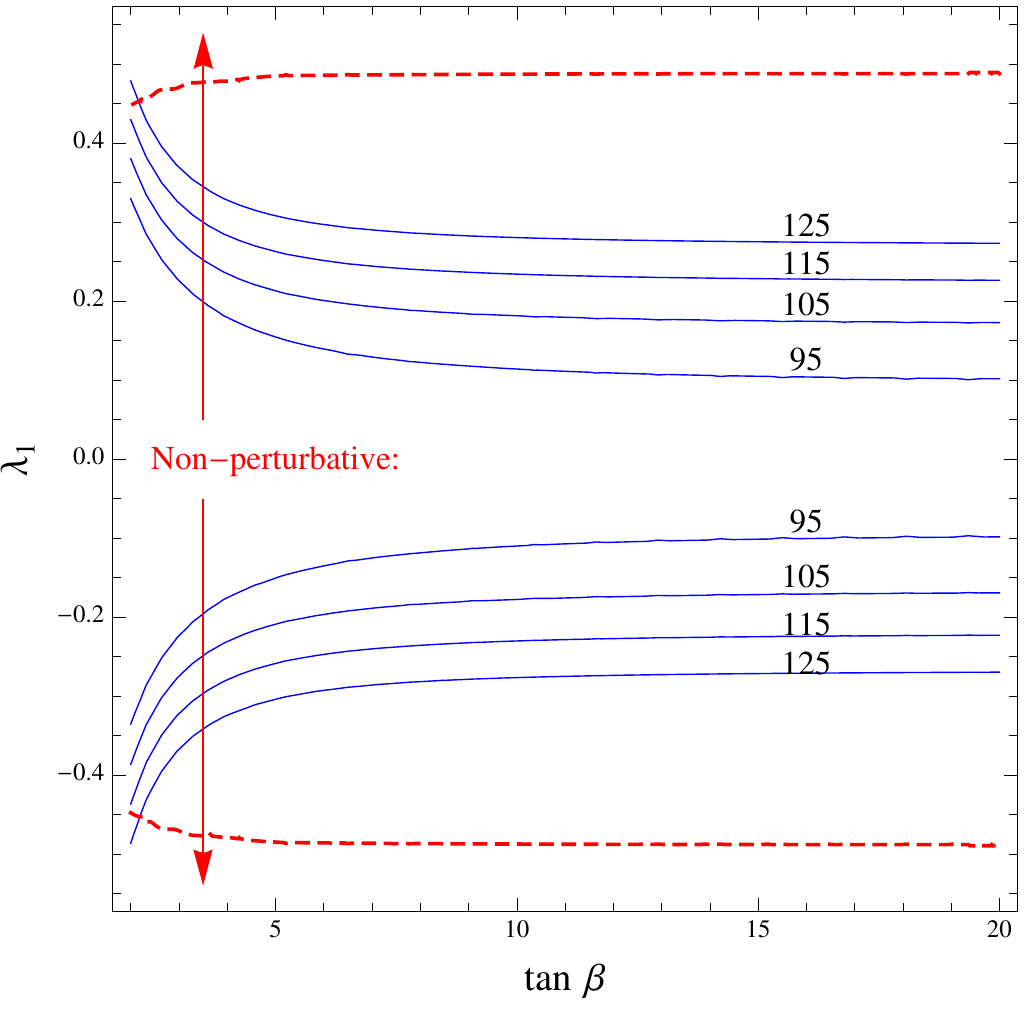}
\caption{Curves of constant tree-level $M_h$ in the $\lambda_1-t_\beta$ plane. The area outside the vertical red dashed lines becomes non-perturbative before the GUT scale under the assumptions discussed in the next section.}
\label{lambda1.tanbeta}
\end{figure}	
Notice that in this model the minimization conditions are more involved and on top of the usual fine-tuning present in the MSSM one has a fine-tuning in the new minimization conditions.
The Higgs mass matrices are given in the Appendix. The Fig.~\ref{mh.v.lambda1} shows the values of the Higgs mass at tree-level versus $\lambda_1$ and a scan over the remaining parameters as follows:
\begin{align*}
	& \tan \beta \supset (2-45); \ |\lambda_2| \supset (-1, 1);
\\
	|\mu|,\ |\mu_{\Delta}| \supset & (100, 500) {\rm{GeV}};  \ \ |v_1|,\ |v_2| \supset (0.5, 1) {\rm{GeV}};
\\
	|b|,\ |b_\Delta| \supset & (50^2, 1000^2) {\rm{GeV}}^2; \ \  |a_1|, \ |a_2| \supset (-1, 1) {\rm{TeV}}.
\end{align*}
Included is the upper bound on the Higgs mass (Eq.~(\ref{ubII})) with $\lambda_2 = 0$ and $\tan \beta = 10$ (in orange). The area outside the vertical red dashed lines is not perturbative up to the GUT scale under the assumptions discussed in the next section. Given the values from the scan, one can use Eqs.~(\ref{MC1}) and (\ref{MC2}) to calculate the soft triplet masses. They are in the following ranges: $m_{\Delta_1} \supset (273, 8900)$ GeV and $m_{\Delta_2} \supset (272, 4630)$ GeV.

Fig.~\ref{mh.v.lambda1} demonstrates that it is possible to achieve a large Higgs mass at tree-level even when $\lambda_1$ is small. In principle, one can even saturate the value of the 125 GeV Higgs mass at tree-level. Here we have assumed that all other Higgs fields are heavier than 300 GeV, the decoupling limit where the lightest Higgs is SM-like.
The contribution from $\lambda_2$ is proportional to $\cos^4 \beta$ and does not play a major role. 
Focusing on the two most relevant parameters, we show in Fig.~\ref{lambda1.tanbeta} a simple tree-level $M_h$ isoplot in the $\lambda_1-\tan \beta$ plane, where the red lines once more delineate the non-perturbative region. One can appreciate that consistency with a 125 GeV Higgs mass is possible at tree-level with all coupling perturbative to the GUT scale.

\begin{figure}[h] 
	\includegraphics[scale=0.7]{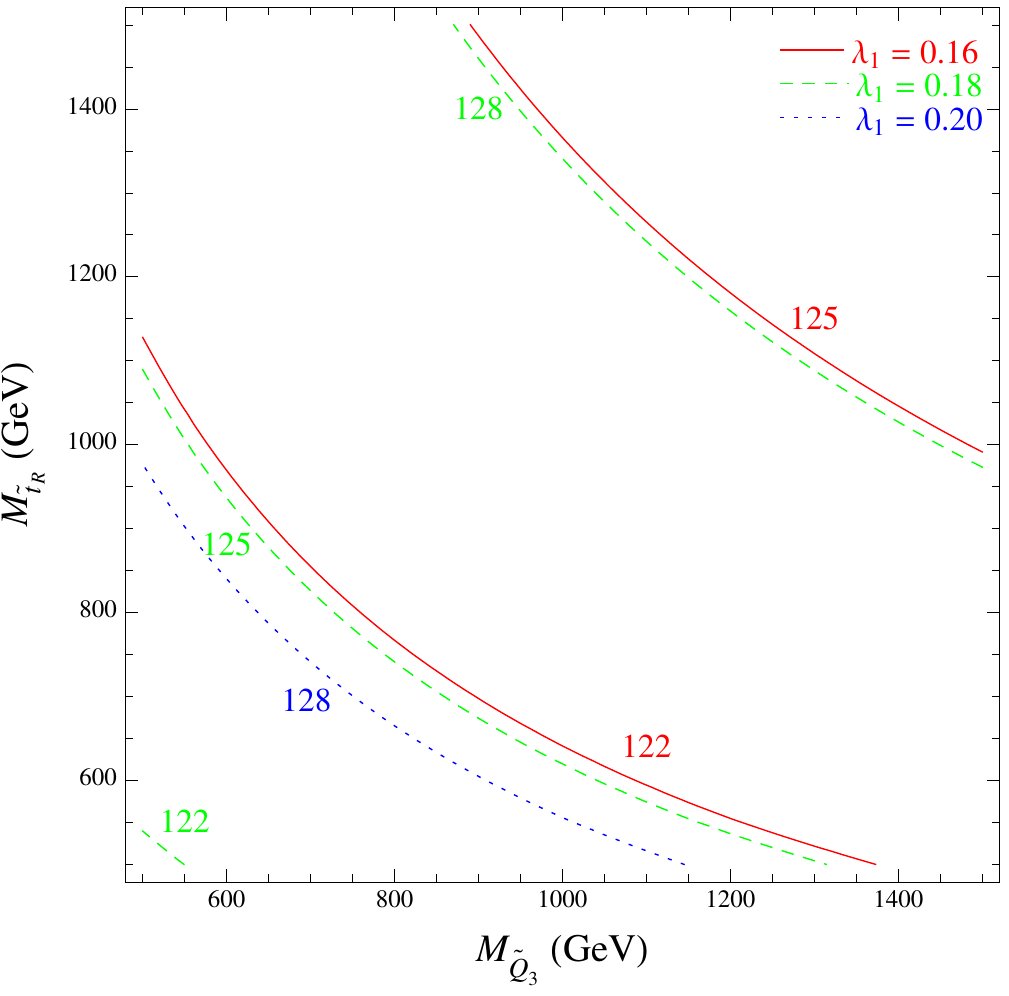}
\caption{Values of constant $M_h$ at the two-loop level in the $M_{\tilde{t}_R}-M_{\tilde{Q}_3}$ plane for zero left-right stop mixing 
and for $\lambda_1 = 0.16, 0.18, 0.20$ in red solid, green dashed and blue dotted curves respectively.}
\label{mtc.mQ3}
\end{figure}

Now, let us discuss the effect of loop corrections to understand how light the stops can be when the tree-level mass is large. 
Curves of constant Higgs mass at the two-loop level with zero left-right mixing in the stop sector are displayed in Fig.~\ref{mtc.mQ3} in the plane of right-handed versus left-handed stop soft masses. Curves for $\lambda_1 = 0.16, 0.18, 0.20$ are solid red, dashed green and dotted blue respectively. Loop calculations were performed using FeynHiggs~\cite{FeynHiggs} 
and we use 1.5 TeV for the gluino mass.
As one can appreciate from Fig.~\ref{mtc.mQ3} the stops can be very light even when the left-right mixing is zero.	
In summary, the type II seesaw mechanism provides the simplest scenario which easily accommodates a 125 GeV Higgs mass for generic values of $\tan \beta$ and is perturbative up to the GUT scale. 
%
\section{IV. Unification and Perturbativity}
\label{pert}
It is well-known that the triplet seesaw fields alone destroy the gauge coupling unification present in the MSSM 
and in order to keep this interesting prediction of the MSSM (assuming the desert) one must add extra fields at the TeV scale. 
In our case one could add fields to complete a $\hat{15}_H$ and $\hat{\overline{15}}_H$ of $SU(5)$. However, 
this possibility is not very appealing since the gauge coupling will not be perturbative up to the GUT scale.
A second possibility was proposed in Ref.~\cite{Calibbi:2009cp}:
\begin{equation}
	\hat D^c \sim (\bar 3, 1, 1/3 ), \ 
	\hat{\bar D}^c \sim (3, 1, - 1/3), \
	\hat \Phi \sim \left(8, 1, 0\right),
\end{equation}
where these fields are called ``magic" fields since they do not complete an $SU(5)$ representation 
with the seesaw triplets but still allow for gauge coupling unification.

The superpotential relevant for the new fields is given by
\begin{eqnarray}
	{\cal{W}}_C &=& \frac{1}{2} \mu_\Phi \tr \hat \Phi^2 \ + \
		\frac{1}{3} \eta \tr \hat \Phi^3 + \zeta \hat D^c \hat \Phi \hat{\bar D}^c + \xi \hat d^c \hat \Phi \hat{\bar D}^c
		\nonumber \\
	&+ & y_{D} \hat Q \hat H_d \hat D^c	+ \mu_{D} \hat D^c \hat{\bar D}^c + \mu_b \hat b^c \hat{\bar D}^c,
\end{eqnarray}
It is important to mention that in this model we have assumed generic mass terms for all fields and the conservation of R-parity. However, it is also possible to introduce a  singlet and a $Z_3$ symmetry (a la NMSSM) in which all SUSY mass terms are generated after symmetry breaking (this was done without the colored fields in see~\cite{T5}). Therefore, we could replace the mass terms by a new coupling and the singlet field as
$$\mu \to \lambda_H \hat{S}, \ \mu_\Phi \to \lambda_\Phi \hat{S}, \  \ \mu_D \to \lambda_D \hat{S},$$
thereby solving the $\mu$-problem. The renormalization group equations (RGEs) are given in the Appendix and were used to derive the red dashed lines in Figs.~\ref{mh.v.lambda1} and \ref{lambda1.tanbeta}, which delineate the non-perturbative regime. While the $\lambda_1$ beta-function receives a large self-contribution, Eq.~(\ref{RGE.lambda1}), the increased size of the gauge couplings help to keep it perturbative. Beyond $\lambda_1 \sim 0.5$, the system becomes non-perturbative due to this large contribution.  The unification of the gauge couplings assuming that all new particles are at the TeV scale are shown in Fig.~\ref{RGE} as well as an example of the running of the Yukawa couplings relevant for the Higgs mass discussion. For the latter plot, Yukawa couplings not shown are assumed to be 0.1 at the TeV scale except for $f_\nu$, which is insignificantly small due to neutrino masses. As one can see, all couplings are perturbative.
\section{V. Phenomenological Aspects}
Here we discuss some of the relevant phenomenological aspects of the new physical Higgses and matter fields:
\begin{itemize}

\item Doubly Charged Higgses: Because of the relatively large VEVs of the triplets (about one GeV) and small value of the triplet-lepton coupling ($f_\nu$), the decay $\delta^{\pm \pm}_i \to W^{\pm} W^{\pm}$ dominates over $\delta^{\pm \pm}_i \to e_j^{\pm} e_k^{\pm}$ assuming that the channel $\delta^{\pm \pm}_i \to W^{\pm} H_i^{\pm}$ is kinematically suppressed. 
It is therefore not possible to search for lepton number violation in this context. The decays of the doubly charged Higgses into W's have 
been investigated in detail in Ref.~\cite{Tao}. The most interesting signals in this case correspond to the pair production of the doubly charged Higgs fields:
\begin{equation}
	pp \  \to \  H^{\pm \pm} H^{\mp \mp} \  \to \  4 j + e^{\pm}_i e^{\pm}_j + E_T^{\rm{miss}}.
\end{equation}
Such states can be identified at the LHC with high luminosity.

The decays of the doubly charged Higgsinos into a lepton and slepton are also suppressed for the same reason 
mentioned above for the doubly charged Higgs into two leptons. Then, the main decays are into a neutralino and doubly charged Higgs or the decay into a W and a chargino. These channels are quite sensitive to the supersymmetric spectrum but are unique for these type of models.

\item Colored Octets: The field $\Phi \sim (8,1,0)$ can decay into two heavy quarks $D^c$ and $\bar{D}^c$ (a heavy quark $\bar D^c$ and a down-type quark) through the Yukawa coupling $\zeta$ ($\xi$)
or into two gluons at the one-loop level through the cubic term $\rm{Tr} \ \Phi^3$. Therefore, pair production of these fields through QCD interactions yields finals states with four jets, where the invariant mass of two jets corresponds to the mass of the colored octet. The octetino has the 
same quantum numbers as the gluino and it can decay into a colored octet and a gluino, giving rise to signals with multileptons and missing energy if we assume R-parity conservation.

\item Heavy Quarks: The new quarks $D^c$ and $\bar{D}^c$ are vector-like and could modify the Higgs decays. These can mix with 
the bottom quarks and suppress the Higgs branching ratio into two bottom quarks. As any colored field, they can be produced with large 
cross sections and decay into a Higgs and a SM quark through the Yukawa coupling, or into a gauge boson and a quark. 
\end{itemize}  
\section{VI. Summary}
In this article we have shown that the fields necessary for the type II seesaw mechanism for neutrino masses can also raise the SM-like Higgs mass up to 125 GeV even at tree-level. This can be accomplished independently of the value of $\tan \beta$ and all couplings remain perturbative up to the GUT scale. While the fields that complete GUT representations with the triplets would lead to Landau poles for the gauge couplings, preservation of gauge coupling unification can be accomplished through so-called magic fields: a vector-like pair of down-type quarks and a color octet. Some of the collider signals of these fields were briefly discussed and a more detailed analysis will be published in the near future.
\\
{\small \textit{Acknowledgments}:  P.F.P. thanks M. Lindner and Mark B. Wise for comments and discussions. S.S. would also like to thank A. Romanino for a discussion.}
\newpage
\begin{widetext}
\appendix	
\section{\large{Appendix}}
\underline{Minimization Conditions}: The full minimization conditions are
\begin{align}
\label{MC.Full}
\begin{split}
	0 & =
		m_{H_d}^2 + \mu^2 + \frac{1}{2} c_{2 \beta} M_Z^2 + t_\beta \, b + c_\beta^2 \lambda_2^2 v^2
		+ \sqrt 2 v_1
		\left(
			\lambda_2 \mu_\Delta - t_\beta \, \lambda_1 \mu
		\right)
	\\
	&
		+\sqrt 2 v_2
		\left(
			a_2 + t_\beta \, \lambda_2 \mu 
		\right)
		+ 2 \lambda_2^2 v_2^2 - \frac{v_2^2-v_1^2}{v^2} M_Z^2,
\end{split}
	\\
\begin{split}
	0 & =
	m_{H_u}^2 + \mu^2 - \frac{1}{2} c_{2 \beta} M_Z^2 + \frac{1}{t_\beta} b
	+ s_\beta^2 \lambda_1^2 v^2
	- \sqrt 2 v_1
	\left(
		a_1 + \frac{1}{t_\beta} \lambda_1 \mu
	\right)
	\\
	&
	-\sqrt 2 v_2
	\left(
		\lambda_1 \mu_\Delta - \frac{1}{t_\beta} \lambda_2 \mu
	\right)
	+ 2 \lambda_1^2 v_1^2 + \frac{v_2^2-v_1^2}{v^2} M_Z^2,
\end{split}
	\\
	\begin{split}
		0 & =
			m_{\Delta_1}^2 + \mu_\Delta^2 + c_{2 \beta} M_Z^2
			+2 s_\beta^2 \lambda_1^2 v^2 +  \frac{v_2}{v_1} b_\Delta
		\\
		&
			+\frac{v^2}{\sqrt 2 v_1}
			\left(
				c_\beta^2 \lambda_2 \mu_\Delta - s_{2 \beta} \lambda_1 \mu - a_1 s_\beta^2
			\right)
			- 2 \frac{v_2^2 - v_1^2}{v^2} M_Z^2,
	\end{split}
	\\
	\begin{split}
		0 & =
			m_{\Delta_2}^2 + \mu_\Delta^2 -  c_{2 \beta} M_Z^2
			+2 c_\beta^2 \lambda_2^2 v^2 +  \frac{v_1}{v_2} b_\Delta
		\\
		&
			+\frac{v^2}{\sqrt 2 v_2}
			\left(
				a_2 c_\beta^2 + s_{2 \beta} \lambda_2 \mu - s_\beta^2 \lambda_1 \mu_\Delta
			\right)
			+ 2 \frac{v_2^2 - v_1^2}{v^2} M_Z^2.
	\end{split}
\end{align}
\underline{Neutral CP-even Higgses}:
In the basis $\Re\left(H_d^0, H_u^0, \delta^0_1, \delta^0_2 \right)$, the CP-even mass matrix is
\begin{align}
\begin{split}
	(M_S^2)_{11} & =
		c_\beta^2 M_Z^2  + 2 \lambda_2^2 v_d^2
		- t_\beta b
		+ \sqrt 2 t_\beta \mu
		\left(
			\lambda_1 v_1- \lambda_2 v_2
		\right),
	\\
	(M_S^2)_{12} & =
		b - \frac{1}{2} s_{2 \beta} M_Z^2 + \sqrt 2 \mu \left(\lambda_2 v_2- \lambda_1 v_1\right),
	\\
	(M_S^2)_{13} & =
		\sqrt 2 v
		\left(
			c_\beta \lambda_2 \mu_\Delta - s_\beta \lambda_1 \mu
		\right)
		+ 2 c_\beta \frac{v_1}{v} M_Z^2,
	\\
	(M_S^2)_{14} & =
		\sqrt 2 v
		\left(
			c_\beta a_2 + s_\beta \lambda_2 \mu
		\right)
		+ 2 c_\beta \frac{v_2}{v}
		\left(
			2 \lambda_2^2 v^2 - M_Z^2
		\right),
	\\
	(M_S^2)_{22} & =
		s_\beta^2 M_Z^2  + 2 \lambda_1^2 v_u^2
		- \frac{1}{t_\beta} b
		+ \frac{\sqrt 2}{t_\beta} \mu
		\left(
			\lambda_1 v_1- \lambda_2 v_2
		\right),
	\\
	(M_S^2)_{23} & =
		-\sqrt 2 v
		\left(
			s_\beta a_1 + c_\beta \lambda_1 \mu
		\right)
		+ 2 s_\beta \frac{v_1}{v}
		\left(
			2 \lambda_1^2 v^2 - M_Z^2
		\right),
	\\
	(M_S^2)_{24} & =
		\sqrt 2 v
		\left(
			c_\beta \lambda_2 \mu - s_\beta \lambda_1 \mu_\Delta
		\right)
		+ 2 s_\beta \frac{v_2}{v} M_Z^2,
	\\
	(M_S^2)_{33} & =
		\frac{v^2}{\sqrt 2 v_1}
		\left(
			s_\beta^2 a_1 + s_{2 \beta} \lambda_1 \mu - c_\beta^2 \lambda_2 \mu_\Delta
		\right)
		-\frac{v_2}{v_1} b_\Delta + 4\frac{v_1^2}{v^2} M_Z^2,
	\\
	(M_S^2)_{34} & =
		b_\Delta - 4  \frac{v_1 v_2}{v^2} M_Z^2,
	\\
	(M_S^2)_{44} & =
		\frac{v^2}{\sqrt 2 v_2}
		\left(
			s_\beta^2 \lambda_1 \mu_\Delta - s_{2\beta} \lambda_2 \mu - c_\beta^2 a_2
		\right)
		- \frac{v_1}{v_2} b_\Delta + 4 \frac{v_2^2}{v^2} M_Z^2.
\end{split}
\end{align}
%
\underline{Neutral CP-odd Higgses}:
In the basis $\Im\left(H_d^0, H_u^0, \delta_1^0, \delta_2^0 \right)$, the CP-odd mass matrix is
\begin{align}
	\notag
	(M_P^2)_{11} & =
		-t_\beta b - \sqrt 2 v_2
		\left(
			2 a_2 + t_\beta \lambda_2 \mu
		\right)
		+\sqrt 2 v_1
		\left(
			t_\beta \lambda_1 \mu - 2 \lambda_2 \mu_\Delta
		\right),
	\\ 	\notag
	(M_P^2)_{12} & =
		-b + \sqrt 2 \mu \left(\lambda_2 v_2 - \lambda_1 v_1 \right),
	\\ 	\notag
	(M_P^2)_{13} & =
		\sqrt 2 v \left( c_\beta \lambda_2 \mu_\Delta - s_\beta \lambda_1 \mu \right),
	\\ 	\notag
	(M_P^2)_{14} & =
		-\sqrt 2 v \left(c_\beta a_2 + s_\beta \lambda_2 \mu\right),	
	\\ 	\notag
	(M_P^2)_{22} & =
		-\frac{1}{t_\beta} b
		+\sqrt 2 v_2
		\left(
			 2 \lambda_1 \mu_\Delta  - \frac{1}{t_\beta} \lambda_2 \mu
		\right)
		+ \sqrt 2 v_1
		\left(
			2 a_1 + \frac{1}{t_\beta} \lambda_1 \mu
		\right),
	\\ 
	(M_P^2)_{23} & =
		\sqrt 2 v \left(s_\beta a_1 + c_\beta \lambda_1 \mu\right),
	\notag
\\
	(M_P^2)_{24} & =
		\sqrt 2 v \left( c_\beta \lambda_2 \mu - s_\beta \lambda_1 \mu_\Delta \right),
	\\
	\notag
	(M_P^2)_{33} & =
		\frac{v^2}{\sqrt 2 v_1}
		\left(
			s_\beta^2 a_1 + s_{2 \beta} \lambda_1 \mu - c_\beta^2 \lambda_2 \mu_\Delta
		\right)
		- \frac{v_2}{v_1}  b_\Delta,
	\\
	\notag
	(M_P^2)_{34} & =
		-b_\Delta,
	\\
	\notag
	(M_P^2)_{44} & =
		\frac{v^2}{\sqrt 2 v_2}
		\left(
			s_\beta^2 \lambda_1 \mu_\Delta - s_{2\beta} \lambda_2 \mu - c_\beta^2 a_2
		\right) - \frac{v_1}{v_2} b_\Delta.
\end{align}
\underline{Charged Higgses}:
In the basis $\left(H_d^-, H_u^{-}, \delta_1^-, \delta_2^{-} \right)$, the charged scalar mass matrix is
\begin{align}
\begin{split}
	(M_{\pm}^2)_{11} & =
		s_\beta^2 M_W^2 - t_\beta b
		-\sqrt 2 v_2
		\left(
			a_2 + t_\beta \lambda_2 \mu
		\right)
		+\sqrt 2 v_1
		\left(
			t_\beta \lambda_1 \mu - \lambda_2 \mu_\Delta
		\right)
		\\
		&
		+ 2 \frac{v_2^2}{v^2}
		\left(
			M_W^2 - \lambda_2^2 v^2
		\right)
		- 2 \frac{v_1^2}{v_2^2} M_W^2,
	\\
	(M_{\pm}^2)_{12} & =
		-b + \frac{1}{2} s_{2 \beta} M_W^2,
	\\
	(M_{\pm}^2)_{13} & =
		v \left(s_\beta \lambda_1 \mu - c_\beta \lambda_2 \mu_\Delta \right) - \sqrt 2 c_\beta \frac{v_1}{v_2} M_W^2,
	\\
	(M_{\pm}^2)_{14} & =
		-v \left(s_\beta \lambda_2 \mu + c_\beta a_2 \right) + c_\beta \frac{\sqrt 2 v_2}{v}
		\left(
			M_W^2 - \lambda_2^2 v^2
		\right),
	\\
	(M_{\pm}^2)_{22} & =
		c_\beta^2 M_W^2 -\frac{1}{t_\beta} b
		+\sqrt 2 v_2
		\left(
			\lambda_1 \mu_\Delta - \frac{1}{t_\beta} \lambda_2 \mu
		\right)
		+\sqrt 2 v_1
		\left(
			a_1 + \frac{1}{t_\beta} \lambda_1 \mu
		\right)
		\\
		&
		+ 2 \frac{v_1^2}{v^2}
		\left(
			M_W^2 - \lambda_1^2 v^2
		\right)
		- 2 \frac{v_2^2}{v^2} M_W^2,		
\\
	(M_{\pm}^2)_{23} & =
		- v \left(s_\beta a_1 + c_\beta \lambda_1 \mu \right)
		+ s_\beta \frac{\sqrt 2 v_1}{v}
		\left(
			\lambda_1^2 v^2 - M_W^2
		\right),
	\\
	(M_{\pm}^2)_{24} & =
		v \left(c_\beta \lambda_2 \mu - s_\beta \lambda_1 \mu_\Delta \right) + \sqrt 2 s_\beta \frac{v_2}{v} M_W^2,
	\\
	(M_{\pm}^2)_{33} & =
		\frac{v^2}{\sqrt  2 v_1}
		\left(
			s_\beta^2 a_1 + s_{2 \beta} \lambda_1 \mu - c_\beta^2 \lambda_2 \mu_\Delta
		\right)
		- b \frac{v_2}{v_1} - s_\beta^2 \lambda_1^2 v^2 - c_{2 \beta} M_W^2 + 2 \frac{v_2^2}{v^2} M_W^2,
	\\
	(M_{\pm}^2)_{34} & =
		b_\Delta - 2 \frac{v_1 v_2}{v^2} M_W^2,
	\\
	(M_{\pm}^2)_{44} & =
		\frac{\sqrt 2 v^2}{v_2}
		\left(
			s_\beta^2 \lambda_1 \mu_\Delta - s_{2\beta} \lambda_2 \mu - c_\beta^2 a_2
		\right)
		- b_\Delta \frac{v_1}{v_2} - c_\beta^2 \lambda_2^2 v^2 + c_{2\beta} M_W^2
		+2\frac{v_1^2}{v^2} M_W^2.
\end{split}
\end{align}
%
\underline{Doubly  Charged Higgses}:
In the basis $\left(\delta_1^{--}, \delta_2^{--} \right)$ the doubly-charged scalar mass matrix is
\begin{align}
\begin{split}
	(M_{\pm \pm}^2)_{11} & =
		\frac{v^2}{\sqrt 2 v_1}
		\left(
			s_\beta^2 a_1 + s_{2\beta} \lambda_1 \mu - c_\beta^2 \lambda_2 \mu_\Delta
		\right)
		-b_\Delta \frac{v_2}{v_1} - 2 s_\beta^2 \lambda_1^2 v^2 - 2 c_{2\beta} M_W^2 + 4 \frac{v_2^2 - v_1^2}{v^2} M_W^2,
	\\
	(M_{\pm \pm}^2)_{12} & =
		b_\Delta,
	\\
	(M_{\pm \pm}^2)_{22} & =
		\frac{v^2}{\sqrt 2 v_2}
		\left(
			s_\beta^2 \lambda_1 \mu_\Delta  - s_{2\beta} \lambda_2 \mu - c_\beta^2 a_2
		\right)
		-b_\Delta \frac{v_1}{v_2} - 2 c_\beta^2 \lambda_2^2 v^2 + 2 c_{2\beta} M_W^2 - 4 \frac{v_2^2 - v_1^2}{v^2} M_W^2.
\end{split}
\end{align}
\underline{Renormalization Group of Equations}:
%
The equations which describe the evolution of the gauge couplings are the following
\begin{eqnarray}
	\alpha_1(M_\text{GUT})^{-1} & = & \alpha_1^{-1}(M_Z) - \frac{1}{2 \pi}
	\left(
		\frac{41}{10} \log \frac{M_\text{GUT}}{M_Z}  + \frac{4}{3} \log \frac{M_\text{GUT}}{M_{\tilde f_{1,2}}}
		+ \frac{2}{3} \log \frac{M_\text{GUT}}{M_{\tilde f_{3}}}+ \frac{1}{10} \log \frac{M_\text{GUT}}{M_A}
	\right)
\nonumber \\
	& & 
		- \frac{1}{2 \pi} \left(
		  \frac{2}{5} \log \frac{M_\text{GUT}}{M_{\tilde H_{u,d}}}
		  + \frac{2}{5} \log \frac{M_\text{GUT}}{M_{D^c}} \ + \  \frac{18}{5} \log \frac{M_\text{GUT}}{M_{\Delta}}
	\right),
\end{eqnarray}
\begin{eqnarray}
	\alpha_2(M_\text{GUT})^{-1} & = & \alpha_2^{-1}(M_Z) - \frac{1}{2 \pi}
	\left(
		-\frac{19}{6} \log \frac{M_\text{GUT}}{M_Z} + \frac{4}{3} \log \frac{M_\text{GUT}}{M_{\tilde W}}
		 + \frac{4}{3} \log \frac{M_\text{GUT}}{M_{\tilde f_{1,2}}}+ \frac{2}{3} \log \frac{M_\text{GUT}}{M_{\tilde f_{3}}}
	\right.
\nonumber \\
	& & \left. \quad \quad \quad \quad \quad \quad \quad \quad \ \ \
		 + \frac{1}{6} \log \frac{M_\text{GUT}}{M_A} + \frac{2}{3} \log \frac{M_\text{GUT}}{M_{\tilde H_{u,d}}} 
		 \ + \  4 \log \frac{M_\text{GUT}}{M_{\Delta}}
	\right),
\end{eqnarray}
\begin{eqnarray}
	\alpha_3(M_\text{GUT})^{-1} & = & \alpha_3^{-1}(M_Z) - \frac{1}{2 \pi}
	\left(
		-7 \log \frac{M_\text{GUT}}{M_Z} + 2 \log \frac{M_\text{GUT}}{M_{\tilde g}}
		 + \frac{4}{3} \log \frac{M_\text{GUT}}{M_{\tilde f_{1,2}}}+ \frac{2}{3} \log \frac{M_\text{GUT}}{M_{\tilde f_{3}}}
	\right)
\nonumber	\\
	& & 
	 - \frac{1}{2 \pi} \left( 3 \log \frac{M_\text{GUT}}{M_\Phi} + 1 \log \frac{M_\text{GUT}}{M_{D^c}}
	\right),
\end{eqnarray}
\begin{figure}[h!]	
	\includegraphics[scale=0.66]{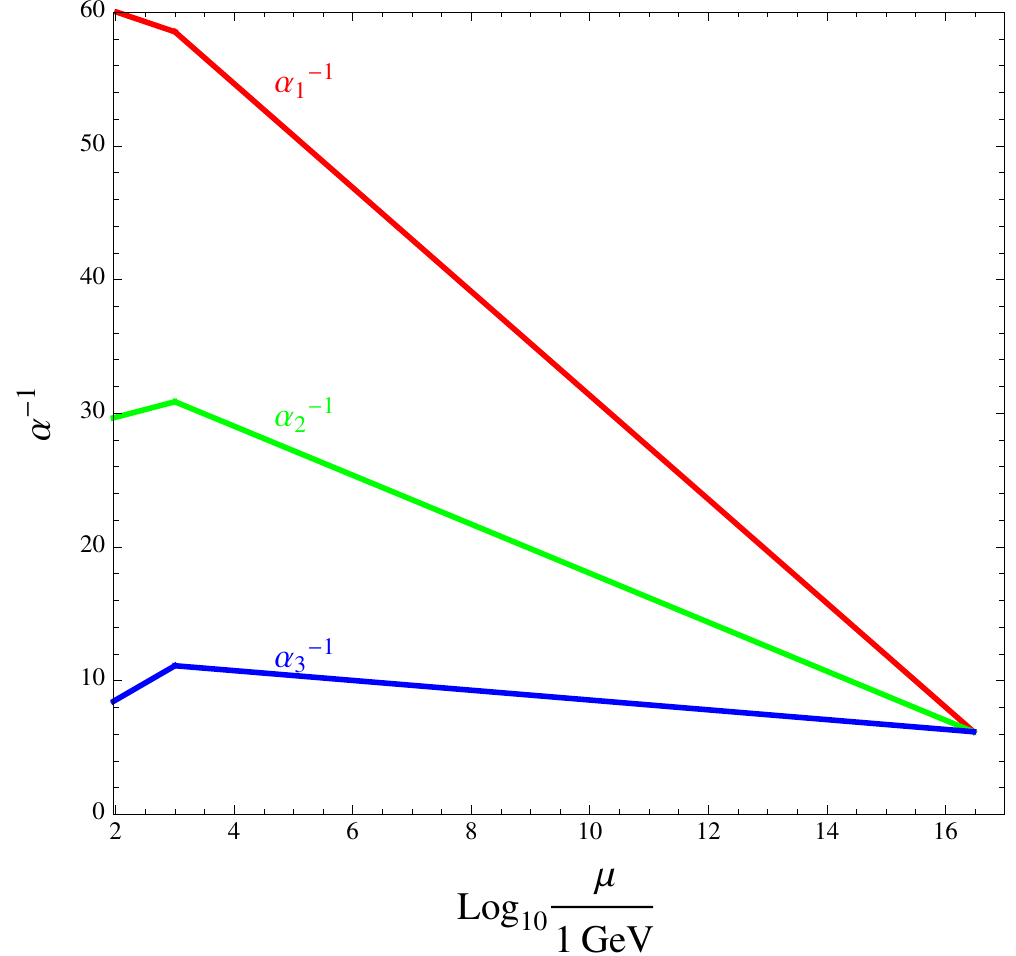}
	\\
	\includegraphics[scale=0.64]{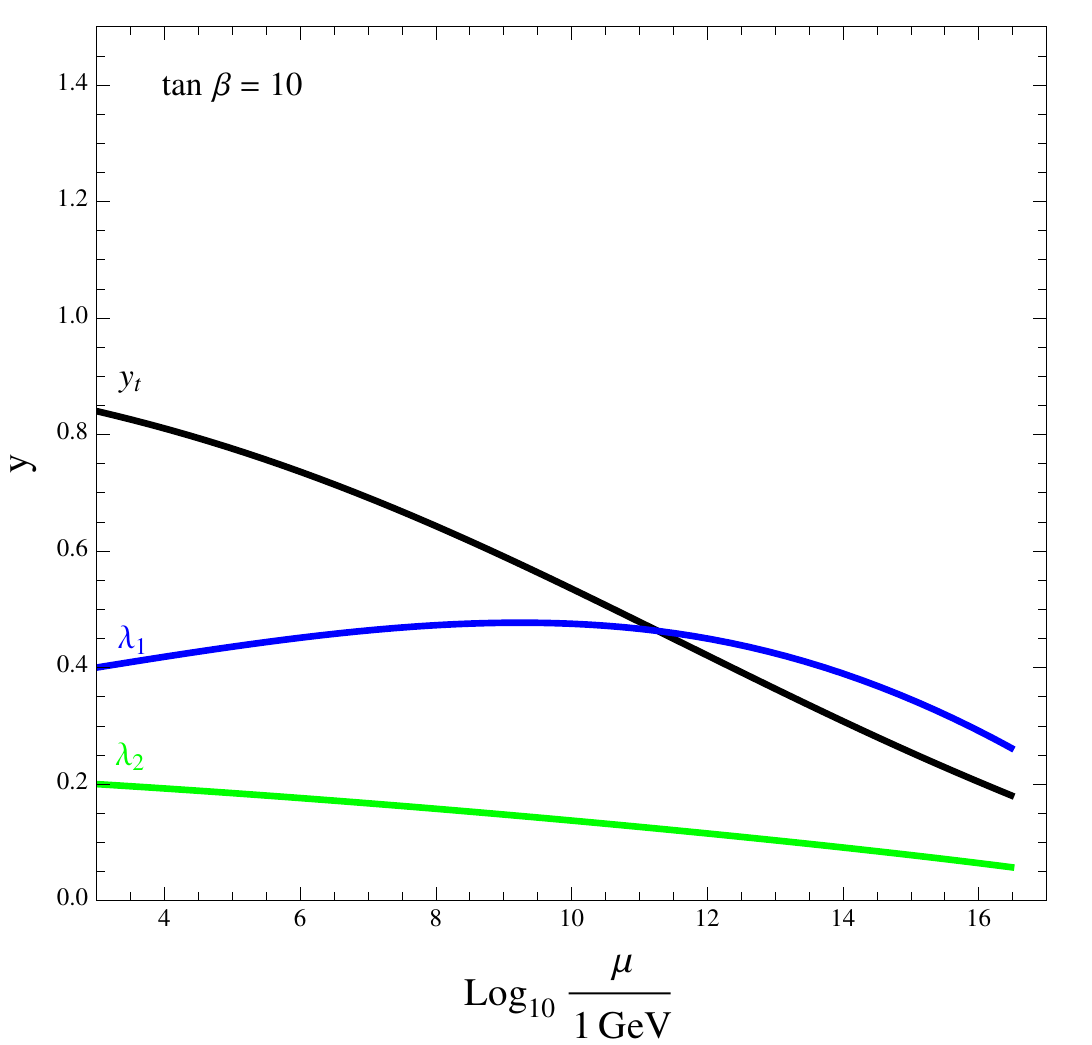}
\caption{Unification of the gauge couplings in the complete model with the colored octet and the new vector-like pair (upper) and the running of the Yukawa couplings relevant to the Higgs mass (lower). Yukawa couplings not shown are 0.1 at the TeV scale except for $f_\nu$, which is insignificantly small due to neutrino masses.}
\label{RGE}
\end{figure}

The RGEs for the Yukawa couplings (here we use the results in Ref.~\cite{Martin:1993zk}) are given by
\begin{align}
	16 \pi^2 \frac{d}{d t} y_t & =
		y_t
		\left(
			6 y_t^2 + y_b^2 + 6 \lambda_1^2 + y_D^2 - \frac{16}{3} g_3^2 - 3 g_2^2 - \frac{13}{15} g_1^2
		\right),
\\
	16 \pi^2 \frac{d}{d t} y_b & =
		y_b
		\left(
			6 y_b^2 + y_t^2+ y_\tau^2 + 6 \lambda_2^2 + 4y_D^2 + \frac{16}{3} \xi^2 - \frac{16}{3} g_3^2 -3 g_2^2 - \frac{7}{15} g_1^2
		\right)
	+ y_D
		\left(
			2 y_b y_D + \frac{16}{3} \zeta \xi
		\right),
\end{align}
\begin{align}
	16 \pi^2 \frac{d}{d t} y_\tau & =
		y_\tau
		\left(
			4 y_\tau^2 + 3 y_b^2 + 6 \lambda_2^2 + 6 f^2_\nu + 3 y_D^2 - 3 g_2^2 - \frac{9}{5} g_1^2
		\right),
	\\
	\label{RGE.lambda1}
	16 \pi^2 \frac{d}{d t} \lambda_1 & =
		\lambda_1
		\left(
			14 \lambda_1^2 + 6 y_t^2 - 7 g_2^2 - \frac{9}{5} g_1^2
		\right),
	\\
	16 \pi^2 \frac{d}{d t} \lambda_2 & =
		\lambda_2
		\left(
			14 \lambda_2^2 + 6 y_b^2 + 2 y_\tau^2 + 2 f^2_\nu  + 6 y_D^2 - 7 g_2^2 - \frac{9}{5} g_1^2
		\right),
	\\
	16 \pi^2 \frac{d}{d t} f_\nu & =
		f_\nu
		\left(
			14 f^2_\nu + 2 y_\tau^2 + 2 \lambda_2^2  - 7 g_2^2 - \frac{9}{5} g_1^2
		\right),
	\\
	16 \pi^2 \frac{d}{d t} \eta & =
		\eta
		\left(
			40 \eta^2 + 3 \zeta^2 + 3 \xi^2 - 18 g_3^2
		\right),
	\\
	16 \pi^2 \frac{d}{d t} \zeta & =
		\zeta
		\left(
			 \frac{35}{3} \zeta^2 + \frac{40}{3} \eta^2 + \frac{19}{3} \xi^2 + 2 y_D^2 - \frac{34}{3} g_3^2 - \frac{4}{15} g_1^2
		\right)
		+ \xi
		\left(
			2 y_b y_D + \frac{16}{3} \zeta \xi
		\right),
\\
	16 \pi^2 \frac{d}{d t} \xi & =
		\xi
		\left(
			\frac{35}{3} \xi^2 + 2 y_b^2 + \frac{40}{3} \eta^2 + \frac{19}{3} \zeta^2 - \frac{34}{3} g_3^2 - \frac{4}{15} g_1^2
		\right)
		+ \zeta
		\left(
			2 y_b y_D + \frac{16}{3} \zeta \xi
		\right),
	\\
		16 \pi^2 \frac{d}{d t} y_D & =
		y_D
		\left(
			6 y_D^2  + y_t^2 + 4 y_b^2+ y_\tau^2 + 6 \lambda_2^2 + \frac{16}{3} \zeta^2
			 - \frac{16}{3} g_3^2 - 3 g_2^2 - \frac{7}{15} g_1^2
		\right)
		+ y_b
		\left(
			2 y_b y_D + \frac{16}{3} \zeta \xi
		\right).
\end{align}

\end{widetext}

\end{document}